\documentclass[twocolumn,aps,superscriptaddress,nofootinbib]{revtex4-1}  % for review and submission

\usepackage[T1]{fontenc}
\usepackage[utf8]{inputenc}
\usepackage{lmodern}
\usepackage{amsmath}
\usepackage{amsfonts}
\usepackage{booktabs}
\usepackage{tabularx}
\usepackage{graphicx}
\usepackage{todonotes}
\usepackage[english]{babel}
\usepackage{hyperref}
\usepackage[caption=false]{subfig}
\usepackage{natbib}

\allowdisplaybreaks %% allows to break up align environments over pages

\newcommand{\md}{\mathrm{d}}

\newcommand{\uu}{{\uparrow \uparrow}}
\newcommand{\dd}{{\downarrow \downarrow}}
\newcommand{\sucker}{\hspace{-15pt}}
\newcommand{\Ntot}{\mathcal{N}}

\begin{document}
\title{Extrinsic curvature in $2$-dimensional Causal Dynamical Triangulation}
\author{Lisa Glaser}
\affiliation{School of Mathematical Sciences, University of Nottingham, University Park, Nottingham, NG7 2RD, UK}
\author{Thomas P. Sotiriou}
\affiliation{School of Mathematical Sciences, University of Nottingham, University Park, Nottingham, NG7 2RD, UK}
\affiliation{School of Physics and Astronomy, University of Nottingham, University Park, Nottingham, NG7 2RD, UK} 
\author{Silke Weinfurtner}
\affiliation{School of Mathematical Sciences, University of Nottingham, University Park, Nottingham, NG7 2RD, UK}

\begin{abstract}
Causal Dynamical Triangulations (CDT) is a non-perturbative quantisation of general relativity. 
Ho\v{r}ava-Lifshitz gravity on the other hand modifies general relativity to allow for perturbative quantisation.
Past work has given rise to the speculation that Ho\v{r}ava-Lifshitz gravity might correspond to the continuum limit of CDT.
In this paper we add another piece to this puzzle by applying the CDT quantisation prescription directly to Ho\v{r}ava-Lifshitz gravity in 2 dimensions.
We derive the continuum Hamiltonian, and we show that it matches exactly the Hamiltonian derived from canonically quantising the Ho\v{r}ava-Lifshitz action.
Unlike the standard CDT case, here the introduction of a foliated lattice does not impose further restriction on the configuration space and, as a result, lattice quantisation does not leave any imprint on continuum physics as expected.
\end{abstract}

\maketitle

\section{Introduction}
Causal Dynamical Triangulations (CDT) is a non-perturbative approach to quantum gravity that 
discretises spacetime into a foliated simplicial manifold. It is an attempt to extend to gravity the lattice methods that have proven very powerful for quantum chromodynamics. CDT has made it possible to numerically explore the path integral over geometries in both $3$ and $4$ dimensions~\cite{ambjorn_nonperturbative_2001,ambjorn_spectral_2005,benedetti_spectral_2009,ambjorn_second-order_2011,ambjorn_effective_2014,ambjorn_recent_2015,benedetti_spacetime_2015}. In  $2$ dimensions the model can be solved analytically~\cite{ambjorn_non-perturbative_1998} and gives rise to a continuum Hamiltonian.

Extending lattice methods to gravity is not straightforward. Instead of calculating field configurations on a fixed lattice,  the lattice itself becomes the object of the dynamics. The presence of a time foliation is crucial. 
The precursor to CDT is the theory of Dynamical Triangulations (DT), where the discretisation 
 is implemented by approximating spacetime through simplicial complexes~\cite{ambjorn_dynamical_1995}, with each $d$ dimensional simplicial complex consisting of $d$-simplices of flat space glued together along their $d-1$ dimensional faces.
In these configurations curvature is concentrated at the $d-2$ dimensional faces of the simplices.
The action on the space of simplicial complexes is the Regge action for discretised spacetimes~\cite{regge_general_1961}.
Simulations of this theory uncovered the existence of two phases, neither of which resembles a continuum spacetime in a suitable limit.
The first is known as the  crumpled phase. Simplices are all glued together as closely as possible and in the limit of infinite size the Hausdorff dimension is infinite as well. The other phase is the branched polymer phase, where the simplices form long chains and the Hausdorff dimension of the resulting space is $2$ \cite{ambjorn_dynamical_1995}.

The solution Ambj\o rn and Loll proposed for this problem was to force the simplicial complex to have a foliated structure~\cite{ambjorn_non-perturbative_1998}.
This gives rise to a unique timelike direction. The length of a timelike edge over the length of  a spacelike edge is a free parameter, $a_t$.
The path integral over these foliated simplicial complexes shows that the resulting geometries are much better behaved.
The $2$-dimensional model can be solved analytically in different ways~\cite{durhuus_spectral_2009,ambjorn_non-perturbative_1998}, which lead to the same result. These approaches have been extended to include matter~\cite{ambjorn_new_2012} or local topology changes~\cite{ambjorn_topology_2008}.

In $3$ and $4$ dimensions analytic methods are no longer fruitful and CDT has been explored through computer simulations.
These have shown that there exists a region in CDT parameter space in which the average Hausdorff dimension of geometries agrees with the dimension of the building blocks, and in which the evolution of spacelike slices follows a mini-superspace action~\cite{ambjorn_semiclassical_2005,benedetti_spacetime_2015}.
This phase has also given rise to the first predictions of a varying spectral dimension~\cite{ambjorn_spectral_2005}, which has been found independently in many other approaches~\cite{lauscher_fractal_2005,horava_spectral_2009,sotiriou_dispersion_2011}(see also Ref.~\cite{carlip_dimensional_2015} for a review and comparison).

Using foliated simplicial complexes might have led to a phase with desirable properties, but introducing a foliation is a thorny issue. Even though the path integral in CDT sums over different foliations it only sums over geometries that actually admit a global foliation. It is thus unclear if one should expect to recover general relativity in the continuum limit or a theory in which all geometries admit a global foliation.

Ho\v{r}ava-Lifshitz (HL) gravity \cite{horava_quantum_2009} is a typical example of a theory with this characteristic. It is a continuum theory with a preferred foliation whose defining symmetry are foliation-preserving diffeomorphisms. Due to the existence of this foliation, one can add higher-order spatial derivatives without increasing the number of time derivatives. This leads to a modification of the propagators at high momenta that renders the theory power-counting renormalizable. In fact, a certain version of HL gravity called {\em projectable} \cite{horava_quantum_2009,sotiriou_phenomenologically_2009} has recently been shown to be renormalizable beyond power counting in 4-dimensions \cite{Barvinsky:2015kil}. In this version the lapse function of the preferred foliation is assumed to be space-independent, which drastically reduces the number of terms in the action and makes the theory tractable. On the other hand, there are serious infrared viability issues concerning projectable 4-dimensional HL gravity \cite{horava_quantum_2009,Sotiriou:2009bx,Koyama:2009hc,Weinfurtner:2010hz,Mukohyama:2010xz,Sotiriou:2010wn} and this suggest that the full non-projectable version \cite{Blas:2009qj} might be phenomenologically preferable.\footnote{Other restricted version of HL gravity exist as well \cite{horava_quantum_2009,Horava:2010zj,Sotiriou:2010wn,Vernieri:2011aa}, but we will not discuss them here.}

It has been shown in Ref.~\cite{horava_spectral_2009} that the spectral dimension in HL gravity exhibits qualitatively the same behaviour as in CDT in 4 dimensions, i.e.~it changes from 4 to 2 in the  ultraviolet. Ref.~\cite{Sotiriou:2011mu} has focused on the simpler case of 3 dimensions, but it has shown that the complete flow of the spectral dimension of (non-projectable) HL gravity from 3 to 2 can reproduce precisely the flow of of the spectral dimension in 3-dimensional CDT.  Interestingly, a certain resemblance can also be found when comparing the Lifshitz phase diagram to the phase diagram of CDT. The measured volume profile of spacelike slices in CDT can be fit with a mini-superspace action  derived from either HL gravity or general relativity~\cite{ambjorn_cdt_2010,benedetti_spacetime_2015}.  These are indications for a connection between  HL gravity and CDT in the continuum limit.

A strong piece of evidence that CDT and HL gravity might be related comes from comparing the Hamiltonians of the $2$d theories. This comparison has been done with projectable HL gravity.  In CDT a continuum Hamiltonian can be derived from the analytic solution of the $2$d theory, while in projectable HL gravity a Hamiltonian can be derived through canonical quantisation. These two Hamiltonians have been compared and found to agree, up to a specific rescaling~\cite{ambjorn_2d_2013}.

The CDT action in 2 dimensions is the discretized version of the Einstein--Hilbert action 
\begin{align}
\label{saction}
S_{2d CDT}= \frac{1}{2 \kappa} \int \md x^2 \sqrt{-g} (R- 2 \Lambda) \to  \lambda \,\Ntot \;,
\end{align}
where $\kappa$ is a dimensionless parameter, $g$ is the determinant of the two dimensional metric $g_{\mu\nu}$, $R$ is the corresponding Ricci scalar and $\Lambda$ the cosmological constant. $\Ntot$ is the total number of simplices and $\lambda$ is the discrete analogue of the cosmological constant.
The action of projectable HL gravity in 2 dimensions is \cite{Sotiriou:2011dr}
\begin{figure*}[!ht]
\subfloat[Transition from down pointing triangle to down pointing triangle]{ \includegraphics[width=0.45\textwidth]{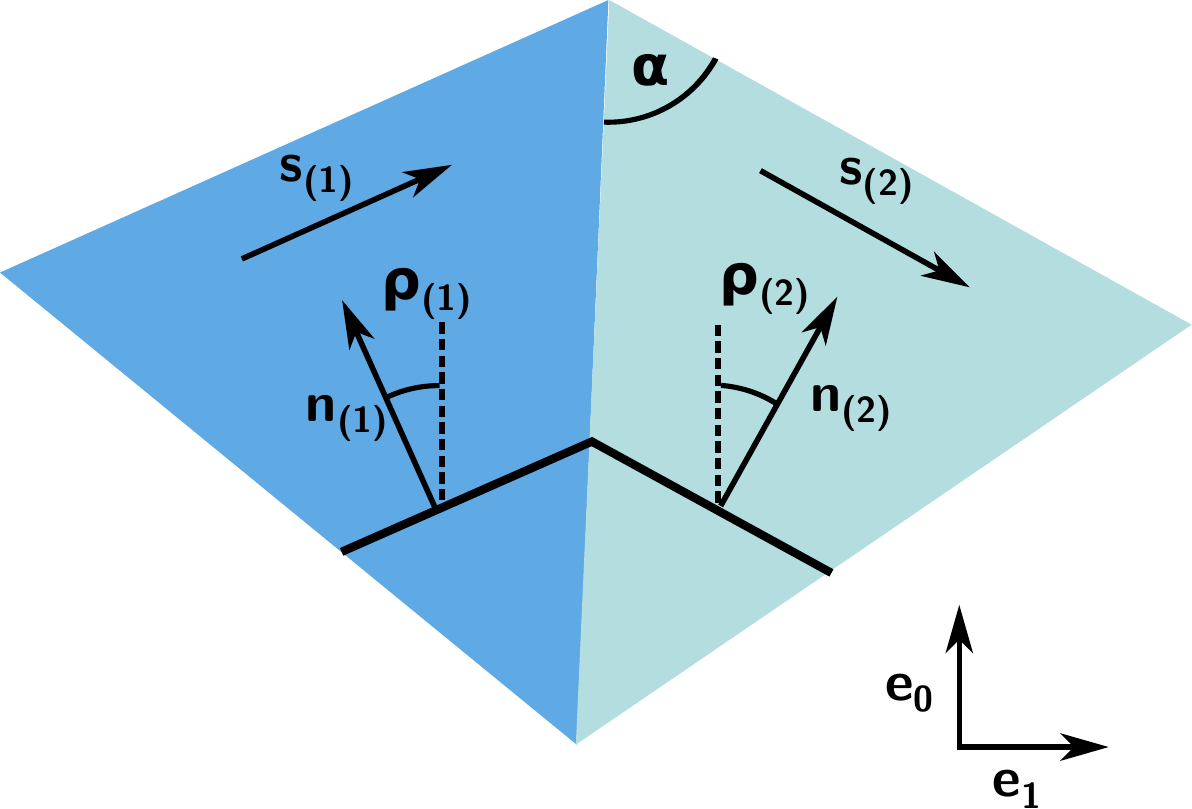}}\hspace{0.05\textwidth}
\subfloat[Transition from up pointing triangle to up pointing triangle]{ \includegraphics[width=0.45\textwidth]{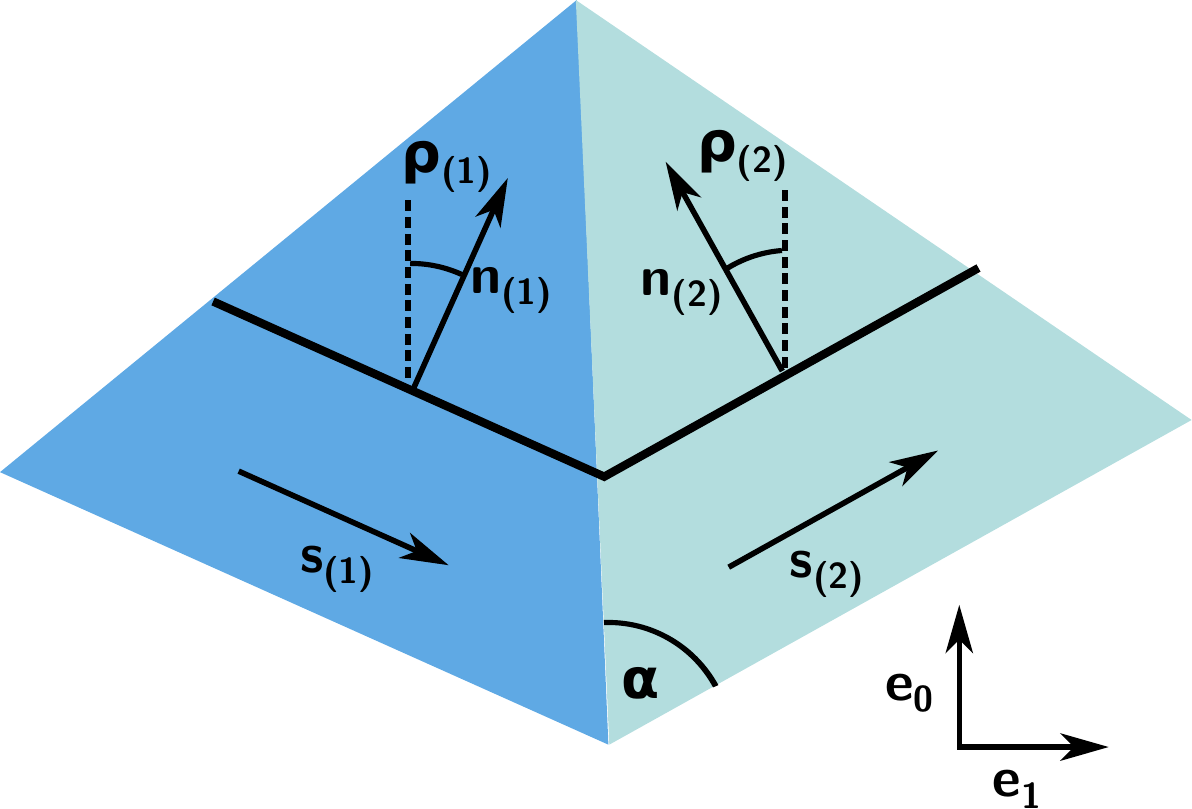}}
\caption{\label{fig:upupdowndown}The angle, $\rho$, between the outward pointing normal vector and the time direction for the coordinate systems at the hinge. The coordinate system is chosen such that $t$ is parallel to the hinge.}
\end{figure*} 
\begin{align}
\label{hlaction}
S_{2d HL} = \frac{1}{2 \kappa} \int \md x \md t N \sqrt{h} [(1-\lambda_{HL}) K^2 - 2 \Lambda]  \;,
\end{align}
where $K$ is the mean curvature of the slices of the preferred foliation, $h$ is the induced metric, and $\lambda_{HL}$ is an extra coupling with respect to GR. 
For $\lambda_{HL}=1$ the only term that survives is the cosmological constant.
This is also the case for the Einstein--Hilbert action (modulo topological consideration) considering that the Ricci scalar is a total divergence in 2 dimensions. 
Though one can in principle absorb the coefficient of $K^2$ in the HL gravity action by suitably redefining the cosmological constant and multiplying the action by a suitable coefficient, this can only be done if no coupling to matter is present and strictly when $\lambda_{HL}\neq 1$.

The fact that the discretised version of the Einstein-Hilbert action and the canonical quantisation of action (\ref{hlaction}) lead to the same Hamiltonian, up to a rescaling, that can be interpreted as fixing $(1-\lambda_{HL})$, is quite intriguing.  
It implies that lattice regularisation of general relativity via CDT does not lead back to general relativity in the continuum limit, but instead to a theory with a preferred foliation. Since CDT restrict the configuration space to that of foliated triangulations, a possible interpretation would be that this restriction leaves its imprint in the continuum limit. 
In this perspective, there seems to be a mismatch between the configuration space and the symmetries of the action in CDT. 
It is thus very tempting to promote the configuration space restriction into an actual symmetry of the (continuum) action, {\em i.e.}~start from a discretisation of an action that is invariant under only foliation-preserving diffeomorphisms, as is  the case for HL gravity.

To this end, instead of applying the CDT prescription to  a discretised version of action \eqref{saction} as in Ref.~\cite{ambjorn_2d_2013}, we apply it to a discretised version of action (\ref{hlaction}). We derive the corresponding continuum Hamiltonian and we compare it with both  the standard CDT continuum Hamiltonian and the Hamiltonian one obtains after canonically quantising HL gravity. 
We show that, for all boundary conditions, we can recover the Hamiltonian for HL gravity, including a free parameter corresponding to $\lambda_{HL}$.
That is, the initial action, and the continuum action one would infer by assigning an action to the continuum Hamiltonian match exactly and share the same continuum symmetries, unlike the case of standard CDT, studied in Ref.~\cite{ambjorn_2d_2013}.

The rest of the paper is organised as follows. In Section~\ref{sec:ext} we find a discrete realisation of the extrinsic curvature squared term for $2$d CDT, which we include in the action in Section~\ref{sec:sum} where we also solve the resulting model analytically.
In Section~\ref{sec:H} we use this analytic solution to derive the Hamiltonian for $2$d CDT with extrinsic curvature terms included, and compare this to the Hamiltonian of projectable $2$d HL gravity.

\section{\label{sec:ext}A discrete extrinsic curvature}

Our  first task is to find an appropriate discretisation for the extrinsic curvature of constant time slices. To this end we will follow the lines of Ref.~\cite{dittrich_counting_2006}, where the extrinsic curvature was used to define trapped surfaces in a triangulation.
It is convenient to actually consider the extrinsic curvature of half-integer time slices $t + \frac{1}{2}$.
This avoids the curvature singularities at the d-2 simplices in the integer $t$ slices.
The extrinsic curvature is concentrated at the joints, or d-1 simplices.

The extrinsic curvature of a spacelike surface $\Sigma$ in a manifold $M$ is given by
\begin{align}
K_{ab}=- h^{c}_{a} \nabla_c n_b
\end{align}
with $n^b$ a unit vector normal to the surface $\Sigma$ and $h_{ac}$ the induced metric on $\Sigma$.
To calculate the extrinsic curvature of the half integer $t$-slices we need the unit  vectors normal to the two pieces of the constant time surface $n^{a}_{(i)}$ and the spacelike unit tangent vectors along the constant time surface $s^{a}_{(i)}$,
\begin{align}
n^{a}_{(i)}&= \text{cosh}\left( \rho_{(i)} \right) \mathbf{e^{a}_0} + \text{sinh}\left( \rho_{(i)} \right) \mathbf{e^{a}_1} \,,\\
s^{a}_{(i)}&= \text{sinh}\left( \rho_{(i)} \right) \mathbf{e^{a}_0} + \text{cosh}\left( \rho_{(i)} \right) \mathbf{e^{a}_1} \;.
\end{align}
$\rho_{(i)}$ is the angle between the normal vector of the $t+\frac{1}{2}$ surface and the d-1 simplex at which the curvature is located.
For the 2d case this is sketched in Figure \ref{fig:upupdowndown}.

The triangles used in CDT are isosceles with a spacelike edge of length $\ell$ at the base and two timelike edges of length $a_t \ell$.  Hence, the angle $\rho$   depends on the base angle $\alpha$,
\begin{align}
\alpha= \text{arccos} \left( \frac{1}{2 a_t} \right) \;.
\end{align}
The relative length parameter $a_t$ lies in the interval $\frac{1}{2}<a_t < \infty$, with the limiting cases clearly being excluded as degenerate, since for $a_t=\frac{1}{2}$ the triangle becomes a spacelike line, and for $a_t= \infty$ it turns into two parallel timelike lines.
This gives us a range for the angle $0 < \alpha < \frac{\pi}{2}$, as we would expect for the base angle of a triangle.
Using this and Fig.~\ref{fig:upupdowndown} we can determine that for the down-down transition $\rho$ is given as
\begin{align}\label{eq:rho}
\rho(x_1) &= \alpha - \frac{\pi}{2} & \rho(x_2) &= -\alpha + \frac{\pi}{2}\,, 
\end{align}
whereas for the up-up transition $\rho$ has the opposite sign.

One  can embed any two triangles into a local Minkowski system such that the kink between them is flat.  The covariant derivative then simplifies to the normal coordinate derivative.
It is straightforward to see from Fig.~\ref{fig:upupdowndown}  that the derivative of the normal vector will diverge as one moves over the kink.
In Ref.~\cite{dittrich_counting_2006} this is resolved by introducing a class of smoothing functions $\delta_\epsilon$ that converge to the delta function as $\epsilon \to 0$.
The angle can then be written as
\begin{align}
\rho(x)= \rho_{(1)} + \Delta \rho \int_{-\epsilon}^{x} \delta_\epsilon(x') \md x'\;,
\end{align}
with $\Delta \rho=\rho_{(2)}-\rho_{(1)}$.

The induced metric can be written as $h_{a}^c=\eta^{c}_{a} +n_{a}n^{c}$, and one can then calculate the extrinsic curvature as
\begin{align}
\label{Kdef}
K^{ab}(x)= - \delta_\epsilon(x) \Delta \rho \;\text{cosh}(\rho(x)) s^a(x) s^b(x)\;.
\end{align}
From this one can calculate the integrated extrinsic curvature scalar as
\begin{align}
K&=\int \lim_{\epsilon \to 0} K(x)\md x= \int \lim_{\epsilon \to 0} K^{ab}(x)\eta_{ab} \md x\\
	&=\int  \delta_\epsilon(x) \Delta \rho \;\text{cosh}(\rho(x))\md x  \;.\label{eq:Kint}
\end{align}
Plugging the $\rho$ values from equation  \eqref{eq:rho} into equation \eqref{eq:Kint} above, we find the integrated extrinsic curvature.
The integrated curvature when passing over down-down, up-up, and up-down transitions are, respectively,
\begin{align}
K_\dd &= (2 \alpha-\pi) \text{cosh}(\alpha - \pi/2) \; ;\\
K_\uu &= -(2 \alpha-\pi) \text{cosh}(-\alpha + \pi/2) \; ;\\
K_{\uparrow \downarrow} &=0 \;.
\end{align}

Here we are not actually interested in the integrated extrinsic curvature itself, but instead in the integral over the extrinsic curvature squared. 
Defining $K^2$ by taking the square of   \eqref{eq:Kint} is problematic due to the presence of the smoothing function $\delta_\epsilon$. This issue can be easily avoided. The smoothing function has been introduced in eq.~(\ref{Kdef}) in order to regularise the curvature on the kink. One can do the same for $K^2$ by defining 
\begin{align}
\label{K^2def}
K^2(x)= - \delta_\epsilon(x) (\Delta \rho)^2 \;\left[\text{cosh}(\rho(x))\right]^2\;.
\end{align}
That can be understood as ``pilling off'' the smoothing function from the definition of $K^{ab}(x)$ in eq.~(\ref{Kdef}) before taking the square and then regularising the result. One can then simply define the integrated squared extrinsic curvature as 
\begin{align}
K^2=\int \lim_{\epsilon \to 0} K^2 (x)\md x \;.
\end{align}
Using this prescription the contribution to the extrinsic curvature squared at each d-1 simplex is
\begin{align}
K_\dd^2 &= (2 \alpha-\pi)^2 \text{cosh}^2(\alpha - \pi/2)\\
K_\uu^2 &= (2 \alpha-\pi)^2 \text{cosh}^2(-\alpha + \pi/2) \;.
\end{align}
Due to the symmetry properties of the hyperbolic cosine these are the same, hence we shall call this term $K ^2$.
Since $0 < \alpha < \pi/2$ one has that $\pi^2 \text{cosh}\left(\pi/2\right)^2 >K^2>0$. We can tune the contribution from each edge by changing the relative edge-length between space and time, but we can not make the contribution vanish or exceed a certain value.

\section{\label{sec:sum}Summing over the simplicial configurations}

We can now include the extrinsic curvature squared term in the simplicial action for a triangulation $T$
\begin{align}
S(T)= \lambda \Ntot + \mu \sum_{\text{transitions}} K^2 \;,
\end{align}
where $\mu$ is the discrete coupling equivalent to $(1-\lambda)/(2 \kappa)$ and transitions refers to all $\uu, \dd $ transitions, since for $\uparrow \downarrow$ transitions the extrinsic curvature vanishes.
Using this discrete action we can  calculate the sum over configurations following the method set out in~\cite{ambjorn_non-perturbative_1998}.

The first step is to calculate the transition function $T^{(s)}_{i j}(g,a,1)$ for a transition from $i$ initial edges to $j$ final edges in one time-step.
In ordinary CDT each configuration from $i$ to $j$ edges has the same weight, since it has the same overall number of triangles. However, in our case the curvature square term adds different weights to different configurations.
After calculating $T^{(s)}_{i j}(g,a,1)$ the next step is to calculate the generating function $\theta^{(s)}(x,y | g,a,1)$.
Switching from the transition function to the generating function is similar to switching from a micro canonical ensemble to a grand canonical ensemble in thermodynamics.
Using the generating function makes many calculations easier, especially taking the continuum limit, in which necessarily $i,j \to \infty$.
It is possible to calculate the generating function for $t$ time-steps by gluing together several generating functions, but for us this step is unnecessary. Instead we will take the continuum limit and expand the generating function to obtain the Hamiltonian of the theory.

Ref.~\cite{di_francesco_integrable_2000} has modified the CDT action by adding a term that contributes at $\uu, \dd$ transitions. The key motivation for adding this terms was to capture the influence of higher curvature corrections. 
In simplicial triangulations the curvature at a given vertex is proportional to $v-6$, where $v$ is the number of triangles adjacent to the vertex.
Hence, in order to construct a term that influences the local curvature they propose to add to the action the terms $|v_1 -3|$ and $|v_2-3|$, where $v_1$ is the number of triangles adjacent to a vertex in the slice above it and $v_2$ is the number of triangles adjacent in the slice below it. Attaching a weight of $a^{|v_1-3|/2+|v_2-3|/2}$ to each vertex is equivalent to attaching a weight of $a$ to each $\uu$ or $\dd$ transition.
The generating function for such a modification has been calculated in Ref.~\cite{di_francesco_integrable_2000}. 
The extra term in our action leads to the same contribution to the discrete path integral as that considered in Ref.~\cite{di_francesco_integrable_2000}. Hence, even though the physical motivation we used to justify this modification of the action is distinct from that used in Ref.~\cite{di_francesco_integrable_2000}, we can nonetheless use the results obtained there.

As we will discuss in more detail latter, the Hamiltonian depends on the boundary conditions and there are more than one options. Ref.~\cite{ambjorn_non-perturbative_1998} applied the closed loop conditions, whereas Ref.~\cite{di_francesco_integrable_2000} solve their model using so called staircase boundary conditions. The latter require that the strip of spacetime has a triangle pointing up on its leftmost edge and a triangle pointing down on its rightmost edge.
This is called a staircase because it resembles one in the dual graph description.
Each of these up/down pointing final triangles has a weight of $\sqrt{g}$ attached.
This is necessary to match to the original result for periodic boundary conditions, as will be explained later.

\begin{figure}[t]
\includegraphics[width=\columnwidth]{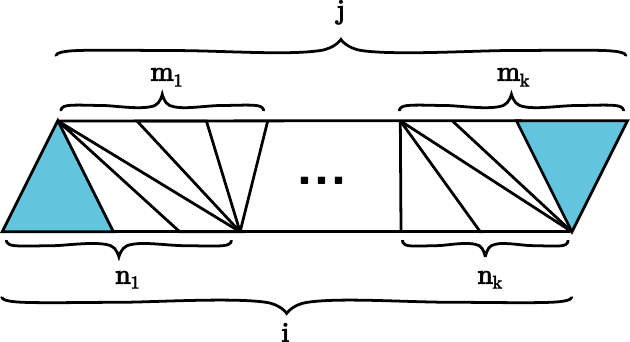}
\caption{\label{fig:Tij}A triangulation going from $i$ initial to $j$ final edges is split into $k$ bunches of $n_r,m_r$ upwards / downwards pointing triangles.}
\end{figure}

In order to write the sum over all triangulations we define $g=Exp(-\lambda)$ and $a=Exp(- \mu K^2)$.
For $\mu = 0$ the extrinsic curvature contribution vanishes and we recover the standard CDT results.
The one time-step transfer matrix connecting $i$ initial to $j$ final edges is given by
\begin{align}
T^{(s)}_{i j}(g,a,1)&= \sum^{\text{min}(i,j)}_{ k=1} \sucker \sum_{\substack{n_r,m_r \\ r=1,2,\dots,k \\ \sum n_r =i \sum m_r=j}} \sucker g^{i+j-1} a^{\sum (n_r-1) + \sum (m_r -1)} \\
	&=  \sum^{\text{min}(i,j)}_{ k=1} \sucker \sum_{\substack{n_r,m_r \\ r=1,2,\dots,k \\ \sum n_r =i \sum m_r=j}} \sucker g^{i+j-1} a^{i-k + j-k} \;.
\end{align}
The $i$ intitial and $j$ final edges can be divided into $k$ bunches of adjacent upwards-pointing triangles and $k$ bunches of adjacent downwards-pointing triangles. We denote the number of triangles in the $r$-th bunch of upwards-pointing triangles as $n_r$ and of the $r$-th bunch of downwards-pointing triangles as $m_r$.
This is illustrated in Fig.~\ref{fig:Tij}.
\begin{figure*}[!t]
\centering
\includegraphics[width=0.9\textwidth]{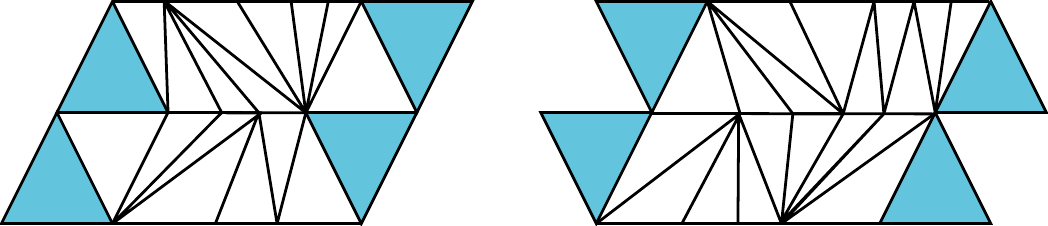}
\caption{\label{fig:stairs}The left figure shows two strips of a triangulation with staircase boundary conditions, while the right side shows two strips of an anti-staircase. These two can be glued together by identifying the blue simplices. }
\end{figure*}
Each composition of $i$ into $k$ terms and $j$ into $k$ terms gives the same weight for a fixed $k$. The sum over the compositions $n_r$ and $m_r$ is then just the number of different compositions, leading to
\begin{align}
T^{(s)}_{ij}(g,a,1)&= g^{i+j-1} a^{i+j}  \sum^{\text{min}(i,j)}_{ k=1} a^{-2k} \binom{i-1}{k-1} \binom{j-1}{k-1}\;.
\end{align}
While each composition into $k$ terms has the same weight, the factor $a$ changes the weight for different $k$, hence leading to a different weighting of the individual geometries than that found in standard CDT.
The next step is  to introduce the generating function, for a single time step
\begin{align}
& \theta^{(s)}(x,y | g,a,1) 	= \sum_{i,j} x^i y^j T^{(s)}_{i,j}(g,a) \label{eq:xysum}\\
					&= \frac{1}{g} \sum_{ k\geq 1} a^{-2k} \sum_{i\geq k} \binom{i-1}{k-1} ( a g x)^i \sum_{j\geq k} \binom{j-1}{k-1} ( a g y)^j \label{eq:sums}\\
					&= \frac{1}{g} \sum_{ k\geq 1} a^{-2k} \frac{ (agx)^k}{(1-agx)^k} \frac{(agy)^k}{(1-agy)^k} \label{eq:eries}\\
					&= \frac{ g x y}{1- ag(x+y) - g^2(1-a^2)xy} \;. \label{eq:onestep}
\end{align}
Diagonalising this single-step generating function and taking it $t$-th power yields a generating function for multiple time steps, $t$. Finally, Di Francesco {\em et al.} take the continuum limit of this function.

We will not repeat this calculation here and instead directly derive a continuum Hamiltonian using equation \eqref{eq:onestep} and the composition rule for $t$-step generating functions.
For this we need to understand the radius of convergence of the sums in eq.~\eqref{eq:sums}. In order to take a continuum limit the coupling constants $g,x,y$ need to be tuned towards their critical values $x_c,y_c,g_c$ which are reached at the radius of convergence.
At these critical values all terms in the sum in eq.~\eqref{eq:sums} make contributions of the same order of magnitude. 

This becomes intuitive when looking at eq.~\eqref{eq:xysum} to determine the values of $x_c,y_c$ at the critical point.
The series converges for $x,y<1$, but only in the limit $x,y \to 1$ do loops of all lengths contribute equally.
Since the continuum limit consists of taking the length of the edges to zero, while taking the number of edges to infinity, we see that only the limit $x,y \to 1$ will lead to loops of non zero macroscopic length.
With $x_c,y_c =1$ fixed we can then determine the radius of convergence of \eqref{eq:onestep}.
We find two possible solutions $g_c=1/(\pm 1 +a)$. Since our solution should smoothly connect to the standard solution for which $a=1, g_c= 1/2$, we conclude that
\begin{align}\label{eq:xcycgc}
x_c&=1 &y_c&=1& g_c&=\frac{1}{1+a} \;.
\end{align}

In addition to the different couplings, the number of geometries included in the sum is also dependent on the boundary conditions imposed.
As already mentioned before, Di Francesco {\em et al.} impose staircase boundary conditions, as these allow one to easily count the possible compositions.
Ordinarily CDT is solved with periodic boundary conditions with or without a marked point.
For our discussion it will be useful to calculate everything for all three of these possible boundary conditions, since we will find that they all find an interpretation in the continuum.

In order to compare the result for staircase boundary conditions with the known results for periodic boundary conditions with one marked point on the in-going boundary Di Francesco {\em et al.}~\cite{di_francesco_integrable_2000} glue the staircase together with an anti-staircase (Fig.~\ref{fig:stairs}). 
An anti-staircase is defined such that the outermost triangles can be glued onto those of the staircase in a way that reproduces the periodic results. See Fig.~\ref{fig:stairs}.

This gluing leads to a two loop correlator with periodic boundary conditions and marked points on both the in-going and outgoing loop.
Attempting to glue the staircase into a single loop would have resulted in a seam with an enforced pattern with down-up down-up (or up-down up-down) pointing triangles. 
The number of configurations with the anti-staircase boundary condition is the same as that of staircase configurations, hence the one step generating functions are identical. 
Gluing the configurations together corresponds to simply multiplying the generating functions, and dividing by $xy$ to remove doubled boundary links.
One then has that
\begin{align}
\theta^{(2)}(x,y | g,a,1)&= \frac{\theta^{(s)}(x,y | g,a)^2}{xy} \nonumber \\ 
					&= \frac{g^2 x y }{(1- a g(x+y) - g^2(1-a^2)xy)^2} \;.
\end{align}
This is the one step generating function for a propagator with a point marked on both the ingoing and outgoing loops.\footnote{The superscript $^{(2)}$ indicates the two marked points, similarly $^{(s)}$ indicates the staircase boundaries,$^{(1)}$ indicates a single marked point and $^{(0)}$ indicates no marked points.}
To convert it to the generating function for the propagator with a marked point only on the incoming loop we unmark the outgoing loop by dividing the amplitude $T^{(2)}_{ij}(g,a,1)$ by  a factor of $i$.
In the generating function this corresponds to calculating
\begin{align}
&\theta^{(1)}(x,y|g,a,1)= \int_{0}^{y} \frac{\md \tilde y}{\tilde y} \theta^{(2)}(x, \tilde y| g,a,1) \;.
\end{align}
We then find
\begin{align}\label{eq:onestepmark}
&\theta^{(1)}(x,y|g,a,1)= \nonumber \\
&\frac{g^2 x y }{(1-a g x)(1- a g (x+y) - g^2(1-a^2)xy)} 
\end{align}
which in the limit $a \to 1$ agrees with the result in \cite{ambjorn_non-perturbative_1998}.
To complete the possible cases we can also calculate the unmarked propagator with periodic boundary conditions, by removing the mark from the incoming loop through $\int_{0}^{x} \md \tilde x/ \tilde x$, and find
\begin{align}\label{eq:onestepnomark}
\theta^{(0)}(x,y|g,a,1)&= \log\left(  \frac{(1-a g y)(1-a g x)}{1- a g(x+y) -g^2 (1-a^2) x y } \right) \;.
\end{align}

\section{\label{sec:H}Deriving a Hamiltonian}
We can derive a Hamiltonian for the development of the loop-loop correlator by combining eq.~\eqref{eq:onestep} with the composition rule for the generating functions. For the generating functions with one marked point, or staircase boundary conditions one has
\begin{align}
&\theta(x,y|g,a,t_1+t_2) =\nonumber \\
& \oint \frac{\md z}{2 \pi i z}  \theta(x,z^{-1}|g,a,t_1) \theta(z,y|g,a,t_2) \;,
\end{align}
where the contour is chosen such that the singularities of $\theta(x,z^{-1}|g,a,t_1)$ are included but  those of  $\theta(z,y|g,a,t_2)$ are not.
This gluing rule is the same for $\theta^{(s)}(x,y|g,a,t)$ and $\theta^{(1)}(x,y|g,a,t)$, since in both cases there is only one consistent way to glue two geometries together along the final / initial boundary.
For $\theta^{(0)}(x,y|g,a,t)$ the composition rule is slightly more complicated \cite{zohren_causal_2009}
\begin{align}
&\theta^{(0)}(x,y|g,a,t_1+t_2) \nonumber \\
							&=\oint \frac{\md z'}{2 \pi i z'^2}  \partial_z \theta^{(0)}(x,z|g,a,t_1)\bigg|_{z=\frac{1}{z'}} \sucker\theta^{(0)}(z',y|g,a,t_2)\;,
\end{align}
taking into account that the final/ initial loops of length $l$ can be consistently glued together in $l$ different ways.
Inserting \eqref{eq:onestep} or \eqref{eq:onestepmark}, or \eqref{eq:onestepnomark} to the suitable one of the two expressions above corresponds to calculating them for $t_1=1$, with $t_2=t-1$. 
This yields
\begin{align}
\theta^{(s)}(x,y|g,a,t) =& \frac{ g x \; \theta^{(s)}\left(\frac{g a + g^2 x(1-a^2)}{1- a g x},y|g,a,t-1\right)
}{ g a + g^2 x (1-a^2)}  \\
\theta^{(1)}(x,y|g,a,t) =& \frac{ g x  \;\theta^{(1)}\left(\frac{g a + g^2 x(1-a^2)}{1- a g x},y|g,a,t-1\right) 
}{ (1-a g x)( a + g x (1-a^2))}\\
\theta^{(0)}(x,y|g,a,t)=& \theta^{(0)}\left(\frac{g a + g^2 x(1-a^2)}{1- a g x},y|g,a,t-1\right) \nonumber\\
						& -\theta^{(0)}\left(g a,y|g,a,t-1\right)\;.
\end{align}

One can calculate the continuum Hamiltonian via an expansion in the lattice spacing.
In the continuum limit the lattice length $\ell$ is taken to zero in such a way that the coupling constants $x,y,g$ are tuned towards their critical points $x_c,y_c,g_c$ which we determined in eq.~\eqref{eq:xcycgc}.
We assume the following scaling around these values
\begin{align}\label{eq:lattice1}
x&= e^{- \ell X}= 1- \ell X +\frac{1}{2} \ell^2 X^2 +O(\ell^3) \\
y&= e^{- \ell Y} = 1- \ell Y +\frac{1}{2} \ell^2 Y^2+O(\ell^3)\\
g&= \frac{1}{a+1} e^{- \ell^2 \Lambda} = \frac{1}{a+1} (1- \ell^2 \Lambda) +O(\ell^4)
\end{align}
with $a$, and hence $a_t$, kept constant.
Since the length of each time step also scales to zero we introduce $t = \tau/\ell$.
The scaling we chose is consistent with that in Ref.~\cite{ambjorn_non-perturbative_1998}, albeit with a slight modification to match the condition $g \to \frac{1}{a+1}$ in the $\ell \to 0$ limit. It  also  matches the scaling chosen in Ref.~\cite{di_francesco_integrable_2000} up to a redefinition of the cosmological constant, $\Lambda \to a \Lambda/2$.
$\Lambda$ is a numerical constant and such a redefinition is legitimate. 
However it will become clear that  the scaling we chose  is preferable when we compare our Hamiltonian to the literature.

We denote the continuum propagators as
\begin{align}
\Theta(X,Y|\Lambda ,a, \tau)&= \lim_{\ell \to 0} \;\ell\; \theta(x,y|g,a,t) 
\end{align}
where $x,y,g,t$ are understood as the functions of $\ell$ defined in \eqref{eq:lattice1} and $\theta$ without a superscript denotes any of the 3 generating functions $\theta^{(s)}$, $\theta^{(1)}$, and $\theta^{(0)}$.
We can then expand $\theta(x,y|g,a,t)$ to first order in $\ell$. This leads to a heat kernel equation
\begin{align}\label{eq:HPDE}
\partial_\tau \Theta(X,Y| \tau , \sqrt{\Lambda},a) &=- H_X \Theta(X,Y| \tau , \sqrt{\Lambda},a) \;, 
\end{align}
with the Hamiltonians
\begin{align}
H^{(s)}_X&= a X + (a X^2 -2 \Lambda )\partial_X \\
H^{(1)}_X&= 2 a X + (a X^2 -2 \Lambda )\partial_X \\
H^{(0)}_X&= (a X^2 -2 \Lambda )\partial_X \;.
\end{align}
From these we can calculate the Hamiltonian acting on $G(L_1,L_2|\tau, \sqrt{\Lambda},a)$ with an inverse Laplace transform,
\begin{align}
H^{(s)}_L&=  -a L \partial_L^2 -a \partial_L + 2 \Lambda L \\
H^{(1)}_L&= -a L \partial_L^2 + 2 \Lambda L \\
H^{(0)}_L&= -a L \partial_L^2  -2 a \partial_L + 2 \Lambda L \;.
\end{align}
It is worth pointing out that the Hamiltonian for the staircase boundary condition is the same as one could derive for an amplitude with two marked points, assuming again that the correct gluing rule is used.

We can now compare these Hamiltonians with the Hamiltonian derived for HL gravity in Ref.~\cite{ambjorn_2d_2013}, where we have reinstated a constant $\zeta=1/(4(1-\lambda_{HL}))$ that is absorbed into the loop length in that paper.
The Hamiltonian from HL gravity actually has three possible forms, depending on the ordering of the operators.
The ordering choice corresponds to the different possible boundary conditions that can be imposed in CDT.
The three possible Hamiltonians are\footnote{The subscripts here are identical to those in Ref.~\cite{ambjorn_2d_2013}, which were chosen to reflect the measure on which the Hamiltonian is hermitian.}
\begin{align}
H_{-1}&= - \zeta L \partial_L^2 + 2 \Lambda L \,,\\
H_{0}&=- \zeta L \partial_L^2 - \zeta \partial_L +2 \Lambda L \,,\\
H_{1}&=- \zeta L \partial_L^2 -2\zeta \partial_L +2 \Lambda L \,.
\end{align}
Identifying $\zeta$ with $a$ there is a complete matching, with $H_{-1}$ matching the Hamiltonian for the single marked loop $H^{(1)}_L$, $H_{0}$ matching the one for the staircase boundary conditions $H^{(s)}_L$, and $H_{1}$ is matching the Hamiltonian for the an unmarked loop $H^{(0)}_L$.

\section{Conclusions}

In this paper we have applied the CDT prescription for quantisation to a discretisation of the action of projectable HL gravity instead of the Einstein-Hilbert action. We have calculated the corresponding continuum Hamiltonians for different boundary conditions and we have shown that they match exactly the Hamiltonians one obtains from the canonical quantisation of HL gravity for different orderings of the operators.

This result is far from surprising and it seems to support the idea that the introduction of a lattice in the quantisation scheme leaves continuum physics unaffected even when the lattice is dynamical. However, this issue is more subtle and this can be better appreciated when our results are interpreted in conjunction with the result of Ref.~\cite{ambjorn_2d_2013}.  It was shown there that  the continuum Hamiltonian for standard 2d CDT agrees with the Hamiltonian for projectable 2d HL gravity up to a rescaling of the loop length $L$ and the cosmological constant $\Lambda$ in HL gravity by a factor $\zeta=1/[4(1-\lambda_{HL})]$. In other words, the starting action did not have a preferred foliation, the final Hamilton did, presumably due to the fact that the configuration space is CDT is restricted to foliated triangulations. Hence, in that case lattice quantisation does seem to leave an imprint on continuum physics.

Combining these two results suggest strongly that if the lattice quantisation scheme is compatible with the symmetries of the original action then it does not affect continuum physics, whereas if the introduction of the lattice introduces further restrictions to the configuration space, then it actually modifies the continuum theory. In standard CDT the requirement that the triangulation be foliated is incompatible between the symmetries of the Einstein-Hilbert action (full diffeomorphisms) and this seems to lead to the generation of the extrinsic curvature terms in the continuum Hamiltonian. 

Considering the process of taking the continuum limit as a form of renormalisation, one can compare this situation with work on the renormalisation group flow in HL gravity.
The large number of couplings of HL gravity in more than 2 dimensions make a complete study challenging, but first studies of part of the parameter space have been done~\cite{contillo_renormalization_2013,rechenberger_functional_2013,dodorico_asymptotic_2014}.
Of particular interest is that they show that the isotropic plane $\lambda_{HL}=1$, which contains GR, is not a fixed plane of the flow~\cite{contillo_renormalization_2013}. Hence, one expects to leave this plane through the generation of symmetry breaking terms.

As already mentioned, the continuum Hamiltonian(s) we derived here are in full agreement with the Hamiltonian(s) of HL gravity, whereas they only agree with the continuum Hamiltonian(s) of standard CDT derived in Ref.~\cite{ambjorn_2d_2013} up to a rescaling of parameters. In the continuum theory this rescaling would correspond to a redefinition of the coupling constant and the cosmological constant, and it could also be seen as a reparametrization of time or the spatial coordinate. Hence, as already discussed in the introduction, it is only allowed without loss of generality if there is no coupling to matter. More generically, it would correspond to a fixing of the HL coupling $\lambda_{HL}$ [to a value different than that corresponding to general relativity]. This is a salient point that certainly deserves further investigation.

Some notes of caution are in order.
Firstly, gravity in $2$ dimensions is significantly different that in higher dimensions, and hence special care needs to be taken in trying to generalize results in $2$d to higher dimension. For example, 2-dimensional general relativity and  HL gravity are topological theories and hence quantisation is trivial. 
In fact,  HL gravity is renormalizable in 2d without any anisotropy between space and time. Secondly, the discretisation of the extrinsic curvature squared term is not unique. Our choice was guided by a balance between physical motivation and solvability. An appropriate discretisation should lead to a good continuum limit and the one we chose manifestly does. However, alternative discretisation scheme do exist~\cite{hamber_higher_1984,ambjorn_quantum_1993}.

Clearly, it would be very interesting to generalise our results to higher dimensions.
While this might not be possible analytically, it can be done numerically.
Some results for simulations of CDT plus higher curvature terms in $2+1$d already exist~\cite{anderson_quantizing_2012}.
It would also be particularly interesting to reexamine the discrete RG flow for CDT in $3$ or $4$ dimensions~\cite{ambjorn_renormalization_2014}, taking into account extrinsic curvature and higher derivative terms.

\begin{acknowledgments}
The authors would like to thank Jan Ambj\o rn and Renate Loll for helpful discussions.
The work leading to this invention has received funding  from the European Research Council under the European Union Seventh Framework Programme (FP7/2007-2013) / ERC Grant Agreement n.~306425 ``Challenging General Relativity''. 
S.W. acknowledges financial support provided under the Royal Society University Research Fellow (UF120112), the Nottingham Advanced Research Fellow (A2RHS2) and the Royal Society Project (RG130377) grants.

\end{acknowledgments}
\bibliography{ref}
\end{document}